\documentclass[a4paper,11pt]{article}
\pdfoutput=1 

\usepackage{jcappub} 

\usepackage[T1]{fontenc} 
\usepackage[makeroom]{cancel}
\usepackage{physics}
\usepackage{empheq}
\usepackage{graphicx}
\usepackage{subcaption}
\usepackage{tabularx}

\usepackage[numbers]{natbib}
 
\newcommand{\ba}{\begin{eqnarray}}
\newcommand{\ea}{\end{eqnarray}}

\def \be {\begin{equation}}
\def \ee {\end{equation}}
\def \bea {\begin{eqnarray}}
\def \eea {\end{eqnarray}}

\newcommand{\eq}[1]{(\ref{#1})}

\title{\boldmath Self-interacting Dark Scalar Spikes around Black Holes via Relativistic Bondi Accretion}

\author[a,b]{Wei-Xiang Feng,}
\author[c,d]{Alessandro Parisi,}
\author[c,e]{Chian-Shu Chen\footnote{Corresponding author},}
\author[f,g]{and Feng-Li Lin\footnote{Corresponding author}}

\affiliation[a]{Department of Physics and Astronomy, University of California, Riverside, CA 92521, USA}
\affiliation[b]{Institute of Physics, Academia Sinica, Taipei 11529, Taiwan}
\affiliation[c]{Department of Physics, Tamkang University, New Taipei 251, Taiwan}
\affiliation[d]{Scuola Normale Superiore, Piazza dei Cavalieri, 7, 56126 Pisa, Italy}
\affiliation[e]{Physics Division, National Center for Theoretical Sciences, Taipei 10617, Taiwan}
\affiliation[f]{Center of Astronomy and Gravitation, National Taiwan Normal University, Taipei 11677, Taiwan}
\affiliation[g]{Department of Physics, National Taiwan Normal University, Taipei 11677, Taiwan}

\emailAdd{wfeng016@ucr.edu}
\emailAdd{alessandro.parisi@sns.it}
\emailAdd{chianshu@gmail.com}
\emailAdd{fengli.lin@gmail.com}

\abstract{We consider the spike mass density profile in a dark halo by self-consistently solving the relativistic Bondi accretion of dark matter onto a non-spining black hole of mass $M$. We assume that the dominant component of the dark matter in the halo is a Standard model gauge-singlet scalar. Its mass $m\simeq 10^{-5}{\rm eV}$ and quartic self-coupling $\lambda\lesssim10^{-19}$ are constrained to be compatible with the properties of galactic dark halos. In the hydrodynamic limit, we find that the accretion rate is bounded from below, $\dot{M}_{\rm min}=96\pi G^2M^2 m^4/\lambda\hbar^3$. Therefore, for $M=10^6~{\rm M}_\odot$ we have $\dot{M}_{\rm min}\simeq1.41\times 10^{-9}~{\rm M}_\odot~{\rm yr}^{-1}$, which is subdominant compared to the Eddington accretion of baryons. The spike density profile $\rho_0(r)$ within the self-gravitating regime cannot be fitted well by a single-power law but a double-power one. Despite that, we can fit $\rho_0(r)$ piecewise and find that $\rho_0(r) \propto r^{-1.20}$ near the sound horizon, $\rho_0(r) \propto r^{-1.00}$ towards the Bondi radius and $\rho_0(r) \propto r^{-1.08}$ for the region in between. This contrasts with more cuspy $\rho_0(r) \propto r^{-1.75}$ for dark matter with  Coulomb-like self-interaction.}

\keywords{accretion, astrophysical fluid dynamics, dark matter theory, massive black holes}

\begin{document} 
\maketitle
\flushbottom

\section{Introduction}
It is well established that dark matter (DM) contributes about one quarter of the total energy of the universe~\cite{Bertone:2004pz}. Many DM models have been proposed, in particular, the introduction of new stable particles. However, the natural properties of DM are still elusive. In the standard $\Lambda\rm{CDM}$ cosmology, DM is assumed to be collisionless. Despite the success of $\Lambda\rm{CDM}$ on large scales, difficulties remain to explain the sub-galactic scales. The core-cusp and missing satellites problems are the discrepancies between the numerical N-body simulations and astrophysical  observations on small scales of structure. The deviations may come from the insufficient implementation of the baryonic processes such as the supernova feedback and photoionization or might be due to the unknown properties of DM. Self-interacting dark matter (SIDM) was first proposed to reconcile these issues with the cross-section per unit mass in the range of $0.45~\rm{cm^2/g} < $ $\sigma/m$ $< 450~\rm{cm^2/g}$ ($m$ denotes the DM mass)~\cite{Spergel:1999mh}. It should be mentioned that some investigations showed that the velocity-independent cross-sections cannot account for the observed ellipticities in clusters~\cite{Yoshida:2000uw,Miralda-Escude:2000tvu}. Furthermore, bullet cluster 1E 0657-56 seems to constrain $\sigma/m$ to be less than $0.7~{\rm cm^2/g}$, and the cross-section of SIDM would be too small to be distinguished from the collisionless DM~\cite{Randall:2008ppe,Zavala:2012us}. While some studies indicate the SIDM with velocity-dependent cross-sections can provide a broader parameter region, $\sigma/m\sim0.1-50~\rm cm^2/g$, to accommodate above puzzles~\cite{Feng:2009hw,vandenAarssen:2012vpm,Tulin:2012wi,Tulin:2013teo}. (see Ref.~\cite{Tulin:2017ara} for a review). The issues of SIDM are still under investigation, in this paper we adopt the constraint in the range of 
\be\label{SIDM-constraint}
0.1~{\rm cm^2/g} < \sigma/m < 1~{\rm cm^2/g} 
\ee
in our study. 

It is generally believed that all large galaxies immersed in dark halos host central supermassive black holes (SMBHs). The dynamics of the presence of a central black hole (BH) will alter the surrounding stellar distribution and generate a density cusp within its radius of influence~\cite{Berezinsky:1992mx,Gondolo:1999ef,Kaplinghat:2000vt,Merritt:2002vj,Merritt:2003qk,Gnedin:2003rj,Gorchtein:2010xa,Lacroix:2015lxa,Lacroix:2016qpq,Shapiro:2016ypb,Alvarez:2020fyo}. It was Peebles~\cite{1972ApJ...178..371P} who first derived a power law for the stellar distribution function based on the scaling argument. He adopted the picture that the stars are bounded in orbits due to the gravitational potential of the central BH and then diffuse into another bound orbit via the star-star gravitational scattering. Bahcall and Wolf obtained the scaling law of steady-state stellar density profile as $n_{\rm star}(r)\propto r^{-7/4}$ in the cusp region by numerically solving the one-dimensional Fokker-Planck equation for spherical clusters and isotropic velocity distribution~\cite{Bahcall:1976aa}. The study is then extended to two-dimensional problems after considering the anisotropy in velocity space and the effect of high eccentricity/low angular momentum orbits (the so-called loss-cone in J-space) considered by Frank and Rees~\cite{Frank:1976uy} and Lightman and Shapiro~\cite{Lightman:1977zz}. Monte-Carlo simulations as well as numerical integration of two-dimensional Fokker-Planck equation were also performed in Refs~\cite{1978ApJ...225..603S,1979ApJ...234..317M,1980ApJ...239..685M,1978ApJ...226.1087C}. Some recent studies taking into account the nonspherical clusters, relativistic corrections, and extreme mass ratio inspirals, etc. can be referred to~\cite{Amaro-Seoane:2012lgq} for a review. On the other hand, for the galactic nuclei one may address the similar question where the DM distributes in the center region of galaxies, and will be entrained towards the massive BH. The redistribution of DM density profile around the massive BH is usually called the central ``spike,'' that might be observed as a point sources of gamma rays ~\cite{Bertone:2005xz} and neutrinos ~\cite{Bertone:2006nq} or through the dephasing of the gravitational waveform induced by DM ~\cite{Kavanagh:2020cfn,Coogan:2021uqv} that can be probed with future interferometers.
Different investigations have shown the adiabatic growth of spike density to be $\rho(r)\propto r^{-1.5}$ and $\rho(r)\propto r^{-\beta}$ with $2.25 \le\beta\le2.5$ under the assumptions of isothermal DM distribution~\cite{Ipser:1987ru} and singular power-law cusp DM distribution~\cite{Quinlan:1994ed,Gondolo:1999ef}, respectively.

In this paper, we aim to provide a self-consistent study of the spike density formed by Bondi accretion~\cite{Bondi:1944jm,Bondi:1952ni} onto a central BH in a typical dark halo dominated by SIDM. The motivation for choosing this setup is as follows. The Bondi accretion is the simplest mechanism to study the accretion near a massive object. Later, one can extend to more realistic accretion dynamics based on the results derived from Bondi accretion. The model of SIDM considered in this paper is a Standard Model gauge singlet self-interacting scalar field with interaction potential $\frac{1}{4!}\lambda|\phi|^4$. From the perspective of particle physics, this model is one of the simplest SIDM models satisfying the current detections~\cite{McDonald:1993ex}. Due to its simplicity and elegance, this SIDM model or its simple extension with interaction potential $\frac{1}{n!}\lambda_n |\phi|^n$ have been widely adopted in the context of cosmology ~\cite{Ford:1986sy,Peebles:1999fz,Peebles:2000yy,Goodman:2000tg,Boyle:2001du,Arbey:2001jj,Riotto:2000kh,Arbey:2003sj,Chavanis:2020rdo}, or the construction of dark boson stars ~\cite{Kaup:1968zz,Ruffini:1969qy,Colpi:1986ye,Jetzer:1991jr,Liddle:1992fmk,Liebling:2012fv,Zhang:2020dfi,Zhang:2020pfh,VasquezFlores:2019eht}. Especially, in ~\cite{Colpi:1986ye} it is for the first time shown that this model can form compact objects due to nonzero self-interaction. Moreover, an EoS of this model is also extracted in the isotropic limit~\cite{Colpi:1986ye}, which can then be adopted for hydrodynamical studies, such as Bondi accretion considered in this paper. Given the exact form of equation of state (EoS), we are able to solve the relativistic Bondi accretion of the flow of this SIDM analytically, and obtain the spike density profile within the range of self-gravitating regime.

Inspired by the early works \cite{Silk:1997xw,Ostriker:1999ee,King:2003ix}, there are recent studies on the spike or cusp profiles around a BH by quite different approaches for various DM models \cite{Shapiro:2014oha,Park:2010yh,Park:2011rf,Richards:2021zbr,Richards:2021upu,Schnauck:2021hlm,Baumgarte:2021thx}. Our work can be seen as an extension along the same line but consider a nontrivial SIDM with a closed form of non-polytropic EoS. This contrasts with the results obtained based on either polytropic EoS or the assumption of velocity-dependent cross-section. The rest of the paper is organized as follows. In Section~\ref{sec:Bondi_accretion_scalar} we study the relativistic Bondi accretion of SIDM in DM halo. We then apply the observational constraints in our model parameters and adopt the fluid description for the spike density profile in Section~\ref{sec:constraint_spike}. We conclude in Section~\ref{sec:conclusions}. Throughout the paper,  $G=\hbar=c=1$ is adopted unless otherwise noted.

\section{\label{sec:Bondi_accretion_scalar}Relativistic Bondi accretion of self-interacting dark scalar}

The resultant profile of the accreting matter due to the Bondi accretion is supposed to depend on the EoS. The simplest EoS is the polytropic type, 
$p=K \rho_0^{\Gamma}$ with $\rho_0$ indicating the mass density, 
which, however, may not be the realistic one for nontrivial DM. In this paper, we will consider a nontrivial but simple DM model which goes beyond the polytropic one. This
is just a massive canonical scalar field with quartic self-coupling, which was first proposed in \cite{Colpi:1986ye} for boson stars, with the following Lagrangian,
\be\label{effective}
\mathcal{L}=-\frac{1}{2} g^{\mu\nu} \phi^{*}_{;\mu} \phi_{;\nu} -V(|\phi|)
\ee
where the scalar potential is given by 
\be\label{potential}
V(|\phi|) = \frac{1}{2}m^2|\phi|^2+\frac{\lambda}{4!}|\phi|^4.
\ee
Here $m$ is the mass of DM mass and $\lambda$ represents the self-coupling strength. This is a simple and viable DM model with discrete symmetry. The resulting cross-section of DM-DM scattering is~\cite{Eby:2015hsq}
\be\label{cross-section}
\sigma=\frac{\lambda^2}{64\pi m^2}\;
\ee
in the non-relativistic limit.
This cross-section should be constrained by \eq{SIDM-constraint} to yield profiles of DM halos consistent with the observed ones. 

To consider the accretion of the above scalar SIDM by a central massive object such as a BH or a compact star of mass $M$, which is described by the Schwarzschild metric 
\be\label{metricb}
{\rm d}s^2=-\left(1-{2M\over r}\right){\rm d}t^2+\left(1- {2M \over r}\right)^{-1}{\rm d}r^2+ r^2({\rm d}\theta^2+\sin^2\theta {\rm d}\phi^2)\;.
\ee 
Instead of directly solving the above Einstein-scalar system in such a background spacetime for the hydrodynamics of the accreting matter, we can consider the regime with $\lambda m_{\rm Pl}^2/m^2\gg1$, where $m_{\rm Pl}$ is the Planck mass, for which  the scalar field only varies on a relatively large length scale $\lambda m_{\rm Pl}^2/m^3\gg 1/m$. Therefore, the stationary scalar field configuration in this regime can be approximated by a perfect fluid for the hydrodynamical study with the following EoS~\cite{Colpi:1986ye}
\be\label{DEoS}
\frac{p}{\rho_B}=\frac{4}{9}\left[\left(1+\frac{3\rho}{4\rho_B}\right)^{1/2}-1\right]^2~{\rm or}\quad~\frac{\rho}{\rho_B}=\frac{3p}{\rho_B}+4\sqrt{\frac{p}{\rho_B}},
\ee
where the free parameter $\rho_B$ is given by
\be\label{B_def} 
\rho_B={3m^4\over 2\lambda}
={3.48 \over \lambda}\left({m \over {\rm GeV}}\right)^4 \times10^{20}~{\rm kg~m^{-3}}.
\ee
The EoS \eq{DEoS} reduces to a condensate fluid $p\propto\rho^2$~\cite{Dalfovo:1999bec} in the non-relativistic limit, $p\ll \rho \ll \rho_B$; while $\rho\simeq 3p\gg \rho_B$ as a radiation fluid in the relativistic limit.
To study the Bondi accretion for the considered SIDM, we can start with either the non-relativistic formulation or the relativistic one. The corresponding sets of the continuity equation and the Euler equation are given in Appendix~\ref{NR_Bondi} and \ref{app:euler}, respectively. However, as shown in Appendix~\ref{NR_Bondi}, the non-relativistic Bondi accretion of the SIDM with EoS \eq{DEoS} gives the relativistic sound speed near the sonic horizon, so that this formulation is not appropriate for our consideration. Therefore, we will adopt the relativistic formulation to consider the Bondi accretion in the following.

Start with the continuity equation, which can be understood as the expression for the constant accretion rate $\dot{M}\equiv {{\rm d} M / {\rm d}t}$, \emph{i.e.},
\be\label{accretion_rate}
\dot{M} \equiv 4\pi r^2 \rho_0 u={\rm constant},
\ee
where $u$ is the negative radial component of 4-velocity of the hydrodynamic flow and $\rho_0$ represents the rest mass density. The mass-energy conservation~\cite{Shapiro:1983du} yields the relation between the total energy density $\rho$ and $\rho_0$, \emph{i.e.}, 
\be\label{1st_law}
\left({\partial\rho \over \partial\rho_0}\right)_{\rm ad}={\rho+p \over \rho_0},
\ee
where the subscript ``ad'' denotes the variation is adiabatic during the accretion process. 
Based on \eq{DEoS} and \eq{1st_law}, one can then derive 
\be\label{rho0a}
\frac{\rho_0}{\rho_B}= {8\over 9} \left(\sqrt{1+{3\rho\over 4\rho_B}} -1\right)\sqrt{3\left(1+2\sqrt{1+{3\rho\over 4 \rho_B}}\right)}
\ee
and the (adiabatic) sound speed square 
\be \label{soundspeed1}
a^2
\equiv\left(\frac{\partial p}{\partial\rho}\right)_{\rm ad}
={1\over 3}\left(1- {1\over \sqrt{1+{3\rho/ 4 \rho_B}}}\right)=\frac{\sqrt{p/\rho_B}}{3\sqrt{p/\rho_B}+2}.
\ee
It is obvious that the existence of a sound barrier at $a={1/ \sqrt{3}}$ when $\rho, p\gg\rho_B$. 
Considering that the sound speed profile $a=a(r)$ is the elementary dynamical quantity characterizing the fluid dynamics, we can invert \eq{rho0a} and \eq{soundspeed1} to express $\rho_0$, $\rho$ and $p$ in terms of $a$. The results are 
\be\label{EoSa} 
\frac{\rho_0}{\rho_B}= {8 a^2 \over 1-3 a^2}\sqrt{1-a^2 \over 1- 3a^2}, 
\quad
{\rho\over\rho_B}={4 a^2 (2-3 a^2) \over (1-3 a^2)^2} 
\quad \mbox{and} \quad 
{p\over\rho_B}={4 a^4 \over (1-3 a^2)^2}\;.
\ee



As for the relativistic Euler equation (see Appendix~\ref{app:euler}), we can integrate it with the help of \eq{1st_law} to the relativistic Bernoulli equation~\cite{Shapiro:1983du}
\be \label{Bernoulli_Eq}
\left({P+\rho \over \rho_0}\right)^2 \left(1+u^2-{2M\over r}\right)=\left(\frac{p_{\infty}+\rho_{\infty}}{\rho_{0,\infty}}\right)^2,
\ee
or equivalently by \eq{EoSa},
\be \label{Bernoulli_Eq_1}
\left({1-a^2 \over 1- 3 a^2}\right)^2 \left(1+u^2-{2M\over r}\right)=\left({1-a_{\infty}^2 \over 1- 3 a_{\infty}^2}\right)^2,
\ee
where the quantity with the subscript $\infty$ denotes its value at $r=\infty$ at which $u$ vanishes. Given an accretion rate and the sound speed at a particular location, we can first solve $u=u(a,r)$ and then turn the Bernoulli equation \eq{Bernoulli_Eq_1} to a profile equation for the sound speed. See Appendix~\ref{app:profile} for details. Besides, the relativistic sonic horizon is better to be defined by the radial location $r_s$ where the \emph{local} Mach number~\cite{Shapiro:1983du}
\be
\mathcal{M}=\frac{u/a}{\sqrt{1-2M/r+u^2}},
\ee
takes the unity value, \emph{i.e.}, $\mathcal{M}_s\equiv \mathcal{M}~|_{r=r_s}=1$. In Appendix~\ref{app:mach}, we sketch how $\mathcal{M}$ characterizes the ``local'' fluid speed over the sound speed. Note that both $\mathcal{M}_s$ and $\mathcal{M}_h\equiv \mathcal{M}~|_{r=r_h}=1/a$ at event horizon $r_h=2 M$ are  independent of $u$. Later on, we will use $\mathcal{M}$ instead of $u$ to characterize the local stream speed of fluid.

\begin{figure}[htbp]%
\center
\includegraphics[width=1\textwidth]{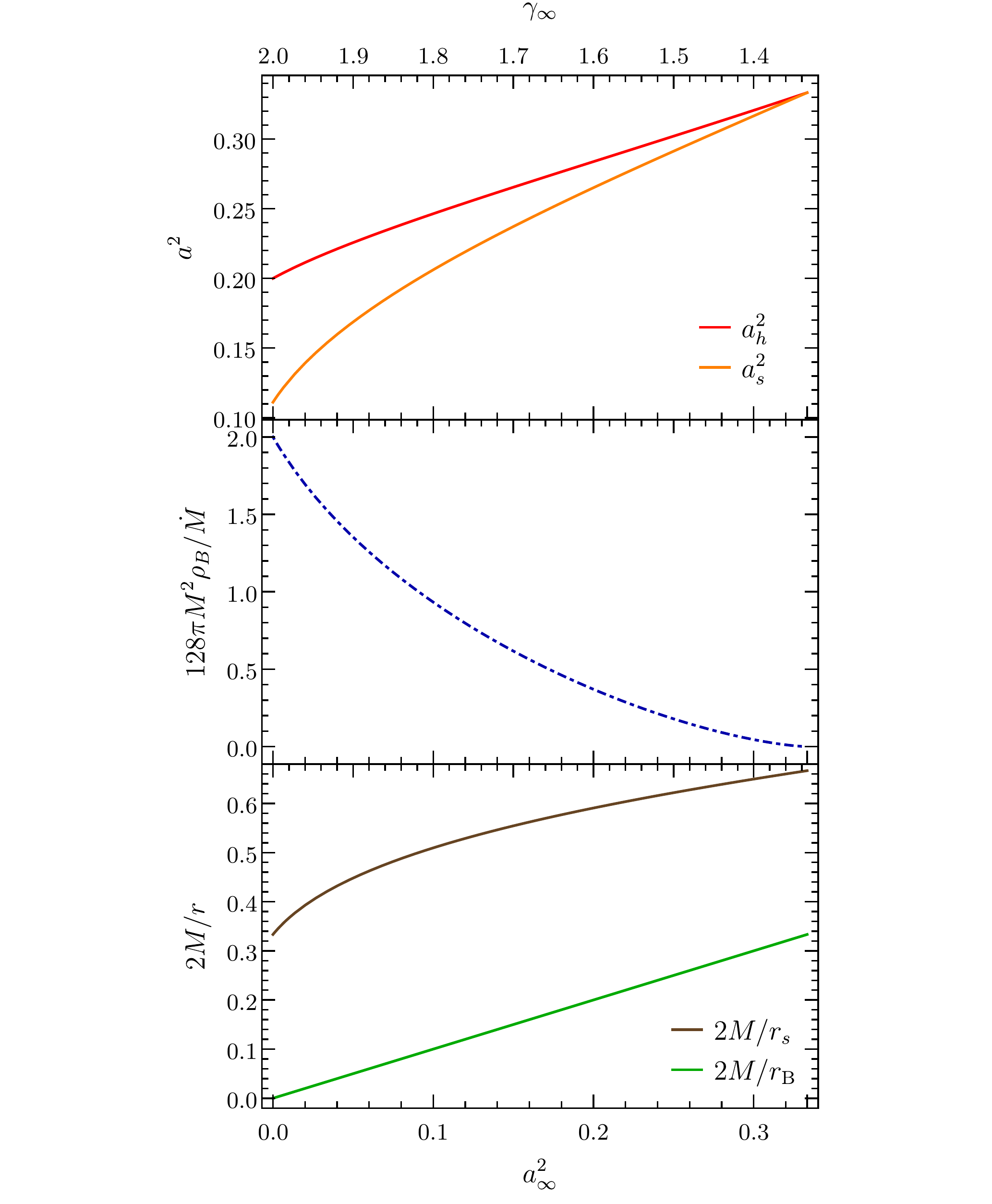}  \caption{Sound speed squares $a_h^2$, $a_s^2$ at different radii (top panel), inverse accretion rate ${128 \pi M^2 \rho_B / \dot{M}}$ (middle panel) and inverse radii $2M/r_s$ and $2M/r_{\rm B}$ (bottom panel), which are the results of Section~\ref{sec:Bondi_accretion_scalar} given $a_\infty^2$ (or adiabatic index $\gamma_\infty$) at infinity. We note that as $\gamma_\infty\rightarrow4/3$, the sound speeds in all range approach the sound barrier $a_\infty^2\rightarrow a_s^2\rightarrow a_h^2\rightarrow1/3$. The accretion rate increases $\dot{M}$ rapidly as the adiabatic index becomes softer, and diverges as $\gamma_\infty\rightarrow4/3$ ($a_\infty^2\rightarrow1/3$). The sound horizon $r_s$ is bounded between $3M$ and $6M$; while the Bondi radius $r_{\rm B}$ between $6M$ and $\infty$. They both decrease as the adiabatic index becomes softer, and vice versa. While the Bondi radius diverges as $\gamma_\infty\rightarrow2$ ($a_\infty^2\rightarrow0$).}
\label{fig:BCplots}
\end{figure}

The profile equation  of sound speed, \emph{i.e.}, \eq{a_r_eq}, does not admit analytical solutions, we will instead solve it numerically in the next section to yield the spike profile of the mass density. Despite that, one can determine the sound speed at sonic horizon $r_s$ and the event horizon $r_h$ (if the center object is a BH) without solving the complicated profile equation. At the sonic horizon, the degenerate feature of the relativistic Euler equation yields  the relations \cite{Shapiro:1983du}
\be \label{sound_horizon}
u_s^2={a_s^2 \over 1+ 3a_s^2} ={M\over 2 r_s}\;.
\ee
This reduces the Bernoulli equation \eq{Bernoulli_Eq_1} which solves the sound speed at sonic horizon, $a_s$ by
\be\label{bcs}
a_s^2={1-3 a^2_{\infty} +\sqrt{1+66 a^2_{\infty}-63 a^4_{\infty}} \over 18 (1-a^2_{\infty})} \;.
\ee
On the other hand, the continuity equation implies $r_h^2\rho_{0h}u_h=r_s^2\rho_{0s}u_s$, which we can solve for the sound speed at the horizon $u_h$ in terms of $a_s$ with the help of \eq{rho0a} and \eq{sound_horizon}. Impose the right-side of \eq{Bernoulli_Eq_1} at the event horizon, then solve it  with the above $u_h=u_h(a_s)$ for the sound speed $a_h$ at $r=r_h$, we obtain  
\be\label{bch}
a_h^2
=\frac{1}{3}-\frac{1}{3}\left[\frac{3}{16}\left({1+3a_s^2\over 1-3a_s^2}\right)^{3/2}\sqrt{1-a_s^2 \over a_s^2}\sqrt{1-3a_\infty^2 \over 1-a_\infty^2}+1\right]^{-1},
\ee
where $a_s$ is given by \eq{bcs}. See the top panel of Fig.~\ref{fig:BCplots} for the numerical profile of \eq{bcs} and \eq{bch}.  Therefore, the physical range of $0\le a_\infty^2<1/3$ 
implies 
\be \label{a_constraints}
{1\over 5} \le a_h^2 < {1\over 3}\;,
\quad \mbox{and}  \quad
{1\over 9} \le  a_s^2 < {1\over 3}\;.
\ee
Remarkably, no matter what value $a_\infty$ takes, the sound speeds at event and sound horizons are always the order of light speed. This justifies the use of the relativistic formalism. 

Furthermore, given the boundary condition, \emph{i.e.}, fixing the value of $a_s$ (or $a_{\infty}$),  then the accretion rate $\dot{M}$ is fixed up to the dimensional parameters $M$ and $\rho_B$.
Specifically by~\eq{rho0a}, \eq{accretion_rate} and \eq{sound_horizon}, one obtains the accretion rate
\be\label{cA_bc}
\dot{M}= 8\pi\left(\frac{1+3a_s^2}{1-3a_s^2}\right)^{3/2}\sqrt{\frac{1-a_s^2}{a_s^2}}M^2\rho_B.
\ee
In an astrophysical system, it is useful to define the Bondi (capture) radius of the accretion range
\be
r_{\rm B}\equiv\frac{2M}{a_\infty^2},
\ee
beyond which the accretion becomes less significant as the particles of the fluid around are no longer gravitationally bound. Consider the two extreme cases: (i) $a_s^2=1/9~(a_\infty=0) \Rightarrow\dot{M}=64\pi M^2\rho_B$, $r_s=6M$, and $r_{\rm B}\rightarrow\infty$; (ii) $a_s^2\rightarrow1/3~(a_\infty\rightarrow1/3)\Rightarrow\dot{M}\rightarrow\infty$, $r_s=3M$, and $r_{\rm B}=6M$. Therefore, one can conclude:
\be\label{accretion_rate1}
64\pi M^2\rho_B\leq\dot{M}<\infty,
\ee
and 
\be
3M\leq r_s\leq6M\leq r_{\rm B}<\infty
\ee
depending on the boundary condition. 
It is interesting to note that the sonic horizon is bound between the smallest circular orbits of massless (photon sphere $3M$) and massive (innermost stable circular orbit $6M$) particles.
We see that the accretion rate increases rapidly as the sound speed $a_{\infty}$ increases, and becomes divergent when approaching the sound barrier, see the middle panel of Fig.~\ref{fig:BCplots}. This contrasts with the result of \cite{Richards:2021zbr} in considering the relativistic Bondi accretion of the matter with polytropic EoS, for which the accretion rate remains finite as $a_{\infty}$ increases.

However, there is a trade-off between the accretion rate ($\dot{M}$) and accretion range ($r_{\rm B}$): \emph{The larger the accretion rate, the smaller the accretion range,} see Fig.~\ref{fig:BCplots} (middle and bottom panels). As long as $r_{\rm B}$ is finite, once the self-interacting bosons are vacuumed up within the region set by $r_B$, the accretion will stop so that it cannot grow indefinitely. Via integration of~\eq{accretion_rate1}, we can obtain the accretion time given the initial and final BH mass,
\be\label{accretion_time1}
\left(\frac{1}{M_i}-\frac{1}{M_f}\right)\geq 64\pi \rho_B\left(t_f-t_i\right).
\ee
Note that the model parameter $\rho_B$ is crucial to the accretion time scale, and it can be determined given $\rho_\infty/\rho_B$ from astrophysical observations.


\section{\label{sec:constraint_spike} Spike profile of the dark halo density around a BH}
In this section, we would like to apply the results derived in the last section to obtain the accreting spike density profile around the central massive object. This is the ideal case of accreting the DM in the halo by a central massive object such as a BH. We will first obtain the numeric windows for the model parameters of the considered SIDM model from the astrophysical observational constraints. With these numerical values of the model parameters, we then numerically solve the profile equation of the sound speed \eq{a_r_eq}, and thus obtain the spike profile of the halo density.

\subsection{Parameters of SIDM for a typical dark halo}
For a virialized system in the fluid description, it is more useful to measure the velocity dispersion square
\be
v_{\rm dis}^2\equiv\frac{3p}{\rho}
=1-\frac{2}{1+\sqrt{1+3\rho/4\rho_B}}
\ee
rather than the sound speed square~\eq{soundspeed1}.
However, they are at the same order of magnitude if $\rho_\infty/\rho_B\ll1$ far away from the Bondi radius. 
Specifically, one can determine the model parameter
\be
\rho_B\simeq\frac{3}{16}\frac{\rho_\infty}{v_{{\rm dis},\infty}^2}
\ee
if $v_{{\rm dis},\infty}^2\ll1$. This is reasonable as the DM halos are generally non-relativistic.
In reality, the dispersion of DM itself is not directly measurable, while it is correlated to that of baryonic matter if the whole system is virialized through gravitational interaction~\cite{Zahid:2018}.
On galactic (cluster) scale, $\rho_\infty\sim10^{-2}~(10^{-3}){\rm~M}_\odot{\rm~pc}^{-3}$ and $v_{{\rm dis},\infty}\sim 10^2~(10^3)~{\rm km~s^{-1}}$~\cite{Tulin:2017ara}, for example, we can determine $\rho_B\sim10^4~(10){\rm~M}_\odot{\rm~pc}^{-3}=6.77\times10^{-16}~(10^{-19})~{\rm kg~m^{-3}}$.
This implies the parameters in our SIDM model to have the relation by utilizing \eq{B_def}, 
\be\label{parameter1}
m\sim10^{-9}~(10^{-10})~\lambda^{1/4}~{\rm GeV}.
\ee
We note that even though the possible $\rho_B$ ranges three orders of magnitude from galaxies to clusters, the mass is still constrained in a narrow window as $m\propto\rho_B^{1/4}$.
Moreover, for typical halos considered above, the lower bound of the accretion rate of \eq{accretion_rate1} yields
\be
\dot{M}_{\rm min}=64 \pi M^2 \rho_B\simeq1.41\times10^{-9}~(10^{-6})~{\rm M}_\odot~{\rm yr}^{-1}
\ee
for BHs of mass $M=10^6~(10^9)~{\rm M}_\odot$ at the central galactic (cluster) mass halos~\cite{Batcheldor:2006ku}, which is subdominant compared to the Eddington accretion~\cite{Salpeter:1964kb} of baryons $\simeq2\times10^{-2}~(10^1)~{\rm M}_\odot~{\rm yr}^{-1}$.

If the SIDM model could also resolve the small-scale structure problem, the constraint of \eq{SIDM-constraint} together with the cross-section in \eq{cross-section} would impose 
\be\label{parameter2}
30\left(\frac{m}{\rm GeV}\right)^{3/2} < \lambda < 90\left(\frac{m}{\rm GeV}\right)^{3/2},
\ee
where the upper (lower) corresponds to the galactic (cluster) scale~\cite{Tulin:2017ara}.
Combining \eq{parameter1} and \eq{parameter2}, we have the self-coupling constant to be constrained in the range of  
\be\label{parameter3}
10^{-22} \lesssim \lambda \lesssim 10^{-19}
\ee
or expressed in terms of scalar mass 
\be\label{parameter4}
10^{-7}~{\rm eV} \lesssim m \lesssim 10^{-5}~{\rm eV},
\ee
which 
overlaps the range of axion DM~\cite{Raffelt:2006cw,Klaer:2017ond}. In addition, the mass window \eq{parameter4} satisfies the Bullet Cluster constraint
$2.92\times 10^{-22}~{\rm eV} < m < 1.10\times10^{-3}~{\rm eV}$~\cite{Chavanis:2020rdo}, and is compatible with 
$4.17~{\rm eV}^4< {m^4/ \lambda} <6.25~{\rm eV}^4$ window\footnote{The values shown here are not exactly the same as in~\cite{Arbey:2003sj} because of the different convention defining the Lagrangian $\mathcal{L}$ and the scalar potential $V(\abs{\phi})$. } 
determined from the rotation curve of the dwarf spiral galaxy, DDO 154~\cite{Arbey:2003sj}.

On the other hand, the accretion time for $M_f\gg M_i$ from~\eq{accretion_time1} results in
\be\label{accretion_time2}
t_f-t_i\equiv\Delta t\lesssim \frac{1}{64\pi \rho_B M_i}=4.8\times10^{12}\left(\frac{\rho_\infty}{\rho_B}\right)\left(\frac{10^{-2}~{\rm M_\odot~pc^{-3}}}{\rho_\infty}\right)\left(\frac{10^6~{\rm M}_\odot}{M_i}\right)~{\rm Gyr},
\ee
which is significantly longer than a Hubble time for a galactic halo of $\rho_\infty/\rho_B\simeq10^{-6}$ with a central BH of initial mass $M_i=10^6~{\rm M}_\odot$. If this SIDM model dominates the main component of the dark halo, it is expected that the accretion is still persisting even for the most massive BHs in our universe. Therefore, it is reasonable to see how the accretion shapes the DM density spikes around BHs.

\subsection{Position-dependent adiabatic index}
Most EoSs for realistic astrophysical consideration are not polytropic, instead they can be approximated as piecewise polytropic functions. We can extend this approximation by introducing the following position-dependent adiabatic index,
\be
\gamma(r) \equiv\left(\frac{\partial\ln p}{\partial\ln \rho_0}\right)_{\rm ad}\;.
\ee
The introduction of $\gamma(r)$ helps to compare with the usual polytropic results whenever ${d\gamma/ dr}$ is small over some region. Besides, $\gamma$ is related to sound speed in a simple manner via \eq{1st_law}, \emph{i.e.}, 
\be \label{soundspeed0}
a^2(r)
\equiv\left(\frac{\partial p}{\partial\rho}\right)_{\rm ad}
=
\frac{p}{\rho+p}\gamma
=1-\frac{\gamma(r)}{2}.
\ee
Applying to EoS \eq{DEoS}, we have
\be
\gamma(r)
=\frac{4}{3}\left(1+{1\over 2\sqrt{1+{3\rho/ 4 \rho_B}}}\right)\;
\ee
which ranges from $2$ ($p\ll \rho \ll \rho_B$) to $4/3$ ($\rho\simeq 3p\gg \rho_B$)\footnote{This lower bound happens when the sound speeds $a_{\infty}$ ( and $a_s$ ) reaches the sound barrier $1/\sqrt{3}$. However, as shown in Fig.~\ref{fig:profiles} for non-relativistic $a_{\infty}\simeq 10^{-7}$, the sound speed can only reach its maximum $1/\sqrt{5}$ at the event horizon so that the lower bound of $\gamma$ is $8/5$ instead of $4/3$.}.
Note that $\rho\rightarrow \rho_0$ when $p\ll \rho\ll \rho_B$ in the non-relativistic limit, thus $p\propto\rho_0^2$ from~\eq{DEoS}, which implies the adiabatic index $\gamma(r)\simeq2$ in sharp contrast to the non-relativistic monatomic ideal gas with the corresponding $\gamma=5/3<2$.
Since the adiabatic index characterizes the \emph{compressibility} of a fluid, the EoS~\eq{DEoS} is \emph{stiffer} due to the \emph{repulsive self-interaction.}

\subsection{Spike profile}
\begin{figure}[htbp]%
\center
\includegraphics[height=1.15\textwidth]{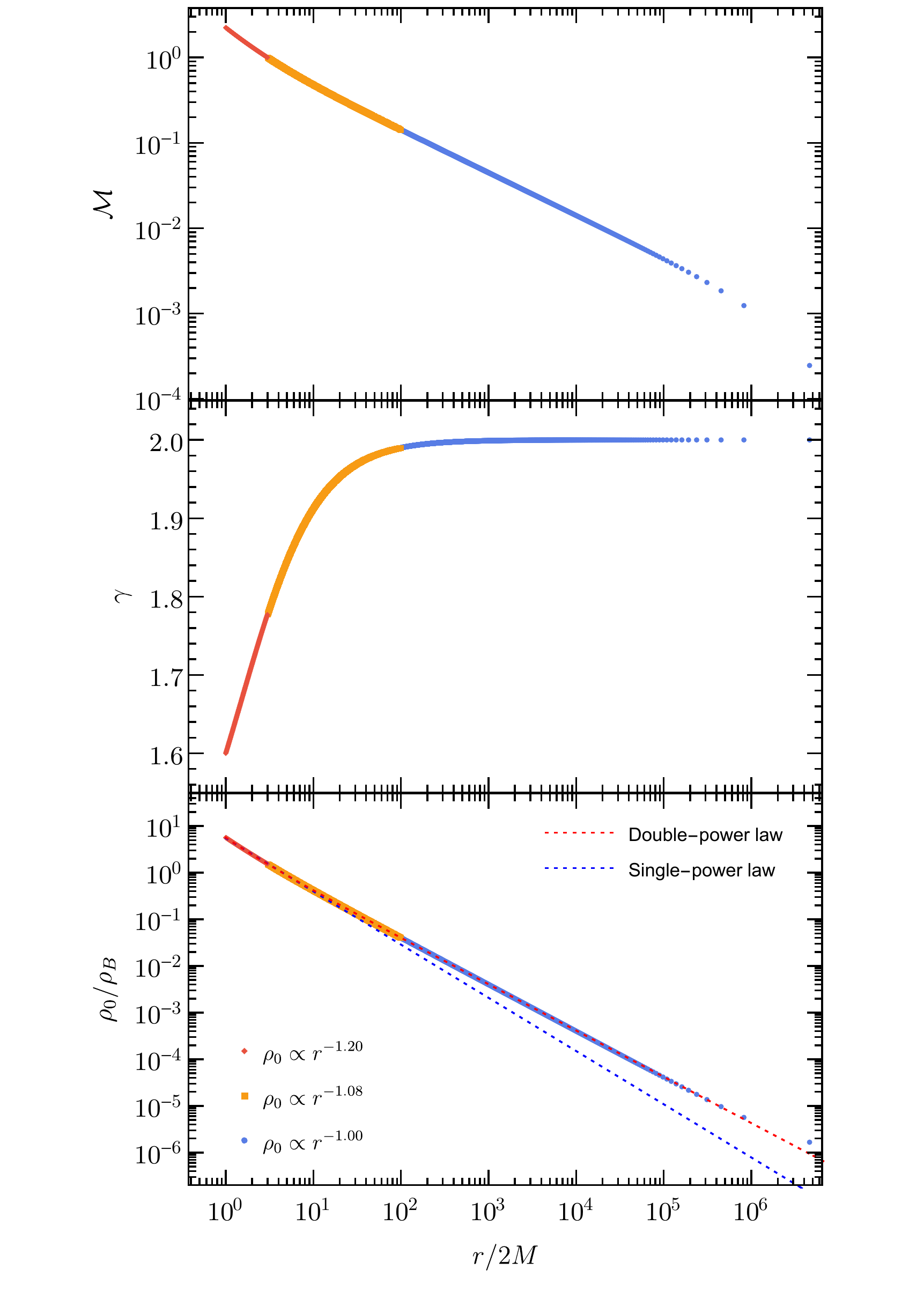}  \caption{Profiles of local Mach number $\mathcal{M}$ (top panel), position-dependent adiabatic index $\gamma$ (middle panel) and mass density $\rho_0$ (bottom panel) for accreting SIDM with EoS \eq{DEoS} around a massive BH. The data are the solutions of \eq{a_r_eq} with $a_\infty^2=10^{-7}$. The profile of $\mathcal{M}$ characterizes the local fluid speed. The profile of $\gamma$ characterizes the local behavior of EoS, which remains stiff ($\gamma\approx2$) in the spike and starts to decrease promptly from $200 M$ towards the central BH ($\gamma\rightarrow\gamma_h=8/5$). The profile of $\rho_0$ shows the spike profile, which can be fitted well by \emph{double-power law} of \eq{doublefit}, but not by the single-power law as shown. As shown, we also fit the mass density profile by single-power law for three regions defined in the main text with small relative error: (1) $\rho_0/\rho_B=5.62(2M/r)^{1.20}$ for near zone (red); (2) $\rho_0/\rho_B= 4.98(2M/r)^{1.08}$ for middle zone (orange); (3) $\rho_0/\rho_B= 4.07(2M/r)^{1.00}$ for far zone (blue).
}
\label{fig:profiles}
\end{figure}

Since the Bondi radius is the range in which the self-gravity effect is relevant for accretion, we shall just consider the spike density profile around the BH within this range. Applying the typical velocity dispersion $v_{{\rm dis},\infty}\sim a_{\infty}$ of galactic halos as discussed, the Bondi radius \be
r_{\rm B}=\frac{2M}{a_\infty^2}\simeq 10^7\times r_h
\simeq\left(\frac{M}{10^6~{\rm M}_\odot}\right)~{\rm pc}\;.
\ee
To obtain the profiles of the sound speed and mass density, we solve the profile equation \eq{a_r_eq} for the sound speed numerically within the aforementioned range of Bondi radius, and then obtain the density profile accordingly via \eq{a_r_eq}. Our results are shown in Fig.~\ref{fig:profiles} for the local Mach number $\mathcal{M}$ (top panel) to characterize the local stream speed of the fluid, the position-dependent adiabatic index $\gamma$ (middle panel) to characterize the sound speed, and the dimensionless spike density profile $\rho_0/\rho_B$ (bottom panel). We see that the local Mach number ranges from unity at the sonic horizon to about $10^{-3}$ near the Bondi radius with a slowly varying $\log{r}$ profile.

As discussed, the position-dependent index $\gamma$ (or equivalently the sound-speed-squared $a^2$) can help to characterize the piecewise feature of EoS. From the variational behavior shown in Fig.~\ref{fig:profiles} this feature is manifest for the EoS we consider, and it can be divided into three regions: (1) the near zone region ranging from the event horizon $r_h$ to the sonic horizon $3 r_h$; (2) the middle zone ranging from the sonic horizon about $3 r_h$ to $100 r_h$; (3) the far zone ranging from $100 r_h$ to the Bondi radius. The far zone is a region of almost constant $\gamma\simeq 2$, and can be thought of as the region with EoS $p\propto \rho_0^2$. On the other hand, $\gamma$ other than the far zone changes more rapidly and ranges from $1.6$ at event horizon to about $2$ around $100 r_h$, so that it cannot be well-approximated by a single polytropic EoS over the near and middle zones. 

Guided by the power-law spike $\rho_0(r)\propto r^{-\beta}$ in the literature \cite{Bahcall:1976aa,Shapiro:1976,Lightman:1977zz} when considering the stellar density around a SMBH, it is tempting to also fit the mass density profile shown in Fig.~\ref{fig:profiles} by the power law. However, the power-law fitting to the \emph{whole} region turns out to be not good. This could be due to the non-polytropic nature of our EoS, \emph{i.e.}, non-constant $\gamma$. Instead, \emph{double-power law} gives a better fit of our model,
\be\label{doublefit}
\rho_0(r)=\left[1.82\left(\frac{2M}{r}\right)^{1.81}+3.84\left(\frac{2M}{r}\right)^{0.99}\right]\rho_B
\ee
with quite small relative error, \emph{i.e.}, $0.558\%$.
The second term $\rho_0(r)\propto r^{-0.99}\approx r^{-1}$ dominates at large radii near the Bondi radius, which can match the Navarro-Frenk-White (NFW)~\cite{Navarro:1995iw} cusp outside the Bondi radius of the BH. 

On the other hand, we can fit some particular region over which $\gamma$ remains constant. In such cases, the result shall take the limiting power-law form obtained in~\cite{Richards:2021zbr} (see (A.8) therein)
\be\label{mass_power}
\rho_0(r)
\propto r^{-{1\over{\Gamma-1}}}\;
\quad{\rm with}\quad
\Gamma>\frac{5}{3}.
\ee
for the accreted matter with polytropic EoS, \emph{i.e.}, $p\propto \rho_0^{\Gamma}$. Indeed, if we fit our mass density profile in the far zone, it is consistent with \eq{mass_power} with $\Gamma=\gamma~\big|_{\rm far~zone}\simeq 2$, \emph{i.e.}, $\rho_0(r)\propto r^{-1}$. Similarly, in the region near the sonic horizon, $\gamma~\big|_{\rm sonic} \simeq 16/9$ so that the mass density fit to \eq{mass_power} with $\Gamma=16/9$, \emph{i.e.}, $\rho_0(r)\propto r^{-1.29}$. Therefore, our results are consistent with the polytropic ones. We have also fit the  mass density in the near, middle, and far zones separately, and the results are: (1) $\rho_0(r)/\rho_B=5.62(r_h/r)^{1.20}$ for near zone (2) $\rho_0(r)/\rho_B= 4.98(r_h/r)^{1.08}$ for middle zone (3) $\rho_0(r)/\rho_B= 4.07(r_h/r)^{1.00}$ for far zone. We see that in all three cases, the spike profiles are less cuspy than the one from the usual Bondi accretion of non-relativistic monatomic fluid with $\Gamma=5/3$, for which the spike behaves as $\rho_0(r)\propto r^{-1.5}$~\cite{Shapiro:1983du}. This is due to the \emph{repulsive} nature of self-interaction of the chosen SIDM model.

In our scenario, the spike profile is formed by accreting the DM, this is different from the Bahcall-Wolf scaling law for the cusp in the number density of stars, \emph{i.e.}, $n_{\rm star}(r) \propto \; r^{-7/4}$ by considering the accretion of  stars into a SMBH at the center of galaxy or cluster. However, we can compare our result with the ones for the accretion of SIDM but different nature of self-interactions. In \cite{Shapiro:2014oha}, the SIDM with velocity-dependent cross-section $\sigma\propto v^{-\alpha}$ is considered for the spike profile, and the resultant spike profile is $\rho_0(r)\propto r^{-(\alpha+3)/4}$. Therefore, the Bahcall-Wolf exponent $\beta=7/4$ can be obtained for $\alpha=4$, which corresponds to the Coulomb-like self-interaction. This contrasts with the results obtained by our velocity-independent cross-section, for which the steepest spike in the near zone is $\rho_0(r)\propto r^{-1.20}$. Thus, the steepest spike profile  from our model is still less cuspy than the Bahcall-Wolf law. Again, this is due to the \emph{repulsive self-interaction} of the chosen SIDM model.

In a broader scenario, the core-expansion phase of SIDM halos can resolve the core-cusp problem of dwarf galaxies~\cite{Sameie:2019zfo} if there is no central BH.
During the core-collapse phase, a central BH
could form directly from DM via gravothermal catastrophe\footnote{Although the SIDM we consider here is stiffer ($\gamma=2$) than the traditional ideal gas ($\gamma=5/3$) in non-relativistic regime~\cite{Feng:2021rst}, it could still trigger the relativistic instability and collapse into a BH if the core density is sufficiently high such that $\gamma$ varies from $2$ towards $4/3$.}~\cite{Balberg:2001qg,Balberg:2002ue,Pollack:2014rja,Feng:2020kxv}. After that, the cusp could still behave as the NFW, $\rho_0(r)\propto r^{-1}$ with the central BH, and match \eq{doublefit}.
 Some comments of possible extensions are made. 
The velocity-independent cross-section $\sigma/m\sim\lambda^2/m^3$ was used in our calculation, therefore, it predicts a constant cross-section universally over all scales and might not be able to describe dwarf halos and clusters simultaneously. In general, for the interaction mediated via a force-carrier might induce velocity-dependent cross-sections and enhance the cross-section by Sommerfeld effect or resonance scattering~\cite{Cirelli:2008pk,Arkani-Hamed:2008hhe}. 

\section{\label{sec:conclusions}Conclusions}

Inspired by obtaining the spike profile by accreting the DM into a central BH inside the dark halo, in this paper we consider the relativistic Bondi accretion of a specific type of DM. This DM model is a massive scalar field with quartic self-interaction, which results in the non-polytropic EoS. This model is well motivated from the perspective of particle physics by its simplicity and elegance. Our work can then be viewed as an interesting interplay between astrophysics and particle physics. 

By solving the relativistic Bondi accretion problem, we find that the corresponding accretion rate, $\dot{M}\geq96\pi G^2M^2 m^4/\lambda\hbar^3$, is bounded from below, and can become divergently large when the initial sound speed approaches the sound barrier. This is quite different from the one for the polytropic type of matter \cite{Richards:2021zbr}.  
Assuming this scalar field dominates the main component of DM around the BH, the shape of the density spike is determined by the model parameters of the DM model, \emph{i.e.}, mass and quartic coupling. Thus, the observation of the spike profile will then put a severe constraint on these model parameters. 
Moreover, due to the repulsive nature of the self-interaction, the spike found in this work is less cuspy than the one predicted by the polytropic type of DM, and also the Bahcall-Wolf power law of the stellar accretion. Specifically, we find the power law varies from $\rho_0(r)\propto r^{-1.2}$ near the BH to $r^{-1}$ towards the Bondi radius for typical dark halos, which are all independent of the mass of the central BH. Overall, the spike density within the self-gravitating region can be well-fitted  by a double-power law of \eq{doublefit}. This is the main prediction of our model, which can be scrutinized by the observational spike data. 

Our formalism can be easily extended to the SIDM with more generic self-coupling. For example, the SIDM with  $V(\phi)={1\over 2} m^2 |\phi|^2 + {\lambda_n \over n!}\abs{\phi}^n$ (for $n>2$) can yield EoS of the form\footnote{private note by Feng-Li Lin.} ${\rho / \rho_*}={n+2 \over n-2}~p/\rho_* + {1\over 2}\left({2n \over n-2}~p/\rho_*\right)^{2/n}$, which becomes $p\propto\rho_0^{n/2}$ in the non-relativistic limit~\cite{Peebles:2000yy}. Following the same procedure of this work, one can solve the spike density due to the relativistic Bondi accretion on to a central BH, although the numerical solution could be more complicated. In the long run, one may pin down the SIDM model by the observation data of spike densities. 
For indirect detection, the self-annihilation of DM can differentiate the small difference in the logarithmic density slope as the annihilation rate depends on the square of the density~\cite{Gondolo:1999ef,Alvarez:2020fyo}, though the annihilation effect could also soften the spike~\cite{Shapiro:2016ypb}. In particular, Event Horizon Telescope can serve as a powerful probe of the DM spikes near BHs~\cite{Lacroix:2016qpq,Alvarez:2020fyo}. However, this deserves further scrutiny for future works.

\acknowledgments
WXF is supported in part by the U.S. Department of Energy under grant No. DE-SC0008541. CSC is supported by Taiwan's Ministry of Science and Technology (MoST) through Grant No.~110-2112-M-032-008-MY3. FLL is supported by Taiwan's Ministry of Science and Technology (MoST) through Grant No.~109-2112-M-003-007-MY3.

\appendix

\section{Inconsistency of non-relativistic Bondi accretion for the SIDM}\label{NR_Bondi}
In this appendix, we show that the non-relativistic Bondi accretion of SIDM obeying \eq{DEoS} is not consistent because it yields relativistic sound speed and fluid speed. 

We start with the non-relativistic continuity equation and Euler equation
\bea 
&& {1\over r^2}{{\rm d} \over {\rm d}r}(r^2 \rho v)=0\;, \label{continuity} 
\\
&& v{{\rm d} v\over {\rm d}r}+{1\over \rho} {{\rm d}p \over {\rm d}r}+{ M  \over r^2}=0\;. \label{Euler}
\eea
Note that in the non-relativistic limit, the rest mass density $\rho_0$ and the total energy density $\rho$ are approximately equal, thus we do not distinguish them here. 
The continuity equation \eq{continuity} implies a constant accretion rate,
\be 
\dot{M}= 4\pi r^2 \rho v.
\ee
Instead of directly solving the Euler equation \eq{Euler}, we rewrite it into the following ``Bondi equation'' \cite{Bondi:1944jm,Bondi:1952ni}: 
\be\label{bondi1}
{1\over 2}\left(1-{a^2 \over v^2}\right) {{\rm d} v^2 \over {\rm d}r}=-{ M \over r^2}\left[1-\left({2 a^2 r \over  M}\right)\right]\;.
\ee
Assuming $a^2(r)$ does not increase too rapidly as $r$ increases, we can see that the RHS will change from positive value to the negative one at the sonic horizon $r=r_s$ with 
\be\label{rs-1}
r_s={ M \over 2a^2(r_s)}.
\ee
This implies that the LHS shall also change the sign accordingly. Since ${{\rm d} v^2 / {\rm d}r}<0$, a physically reasonable initial condition \cite{Bondi:1944jm,Bondi:1952ni} is as follows:
\be
v^2 \rightarrow 0 \quad \textrm{ as } \quad r \rightarrow \infty \qquad \textrm{ and } \qquad v^2(r_s)=a^2_{s}:=a^2(r_s).
\ee
This condition implies that the fluid starts as subsonic fluid at large $r$ and then turns into the supersonic one after crossing the sonic horizon as $r$ decreases.  Subject to the above initial condition, we will now solve the Euler equation \eq{Euler} based on the EoS \eq{DEoS} for SIDM. Therefore, a relativistic formulation of Bondi accretion is needed for the SIDM with EoS given by \eq{DEoS}. 

By the definition of sound speed $a^2=\left({\partial p / \partial \rho}\right)_{\rm ad}$, the EoS \eq{DEoS} can be put into the following form:
\be\label{sounds-2}
{\rho\over \rho_B}={4 a^2 (2- 3a^2)\over (1-3 a^2)^2 }  \qquad \textrm{ or } \qquad \sqrt{p\over \rho_B}={2 a^2\over 1-3 a^2}\;.
\ee
The second expression implies $a^2\le 1/3$ which is the so-called sound barrier. Integrating \eq{Euler} and using \eq{sounds-2} we obtain 
\be
{v^2 \over 2}+ {2\over 3} \ln \left({2-3 a^2 \over 1-3 a^2} \right)-{M\over r}={2\over 3} \ln \left({2-3 a_{\infty}^2 \over 1-3 a_{\infty}^2 } \right)\;.
\ee
where $a_{\infty}=a(\infty)$.  At the sonic horizon $r=r_s$, \emph{i.e.}, \eq{rs-1}, $v^2(r_s)=a_{s}^2$. We can then use this equation to relate the sound speed at sound horizon to the asymptotic sound speed $0\le a_{\infty}\le 1/\sqrt{3}$, and find that 
\be\label{bound-b}
0.38 c \lesssim a_s  \le 0.58 c
\ee
where we have recovered the speed of light $c$. We see that the sound speed at the sound horizon is of the order of $c$, which implies that the non-relativistic formulation of Bondi accretion is not consistent, at least near the sound horizon region.

\section{Relativistic Euler equation in Schwarzschild spacetime}\label{app:euler}
By the conservation law of spherically symmetric stationary inflow in Schwarzschild background, one obtains the relativistic Euler equation~\cite{Shapiro:1983du}:
\begin{equation}
\begin{aligned}
&\frac{u'}{u}+\frac{\rho_0'}{\rho_0}=-\frac{2}{r}~,\\
&uu'+\left(1-\frac{2M}{r}+u^2\right)\frac{a^2}{\rho_0}\rho_0'=-\frac{M}{r^2},
\end{aligned}
\end{equation}
where ``prime'' denotes the derivative with respect to the radial coordinate.
The solution is
\begin{equation}
u'=\frac{\Delta_1}{\Delta}
\quad{\rm and}\quad
\rho_0'=\frac{\Delta_2}{\Delta},
\end{equation}
where
\[
\Delta\equiv\frac{1}{\rho_0 u}\left[\left(1-\frac{2M}{r}+u^2\right)a^2-u^2\right],
\]
\begin{equation}
\Delta_1\equiv\frac{1}{\rho_0}\left[\frac{M}{r^2}-\frac{2a^2}{r}\left(1-\frac{2M}{r}+u^2\right)\right]
\quad{\rm and}\quad
\Delta_2\equiv\frac{1}{u}\left(\frac{2u^2}{r}-\frac{M}{r^2}\right).
\end{equation}
\paragraph{Remark}
(i) $\Delta>0$ as $r\rightarrow\infty$, $u\rightarrow0$ and $a\rightarrow a_\infty$ (subsonic $u^2<a^2$)
(ii) $\Delta=(u/\rho_0)(a^2-1)<0$ at $r=2M$ as $a<1$.
(iii) $\Delta$ must pass through zero at \emph{some critical point} $r=r_s$ outside $r=2M$. To avoid the singularities in the stationary flow, one demands $\Delta=\Delta_1=\Delta_2=0$ at $r=r_s$. Thus, at this \emph{critical point} (sound horizon), one derives~\eq{sound_horizon}.

\section{Deriving profile equation for sound speed}\label{app:profile}
We determine the spatial profile of the Bondi accretion from the relativistic Bernoulli equation \eq {Bernoulli_Eq} given the boundary condition to fix 
\be
c_B \equiv \left(\frac{p_{\infty}+\rho_{\infty}}{\rho_{0,\infty}}\right)^2.
\ee
We aim to solve \eq{Bernoulli_Eq} for the sound speed profile. However, there is still $u$ in the LHS of \eq {Bernoulli_Eq}. We can use \eq{accretion_rate} and \eq{rho0a} to express $u=u(a,\bar{r})$ and the result is
\be\label{u_a}
u={ c_A  (1-3 a^2)^{3/2} \over a^2 \bar{r}^2 \sqrt{1-a^2} }
\ee
where
\be
c_A \equiv {\dot{M} \over 128 \pi M^2 \rho_B}\;,
\ee
and we have introduced the radius coordinate $\bar{r}=r/2M$ in the unit of Schwarzschild radius. 

With \eq{u_a}, we can turn \eq{Bernoulli_Eq} into the  following to solve for $x \equiv a^2$:
\be\label{a_r_eq}
(27 c_A^2 - \bar{r}^3 +\bar{r}^4 -3 c_B \bar{r}^4) x^3 - (27  c_A^2 - \bar{r}^3 + \bar{r}^4 - c_B \bar{r}^4) x^2 + 9 c_A^2 x - c_A^2 =0\;.
\ee
Once the boundary condition is specified, the profile of sound speed can be obtained and, in turn, the density profile around the central hole.
We need to specify the physical boundary conditions, such as $a_\infty$, of the dark halo to fix $c_A$, $c_B$ and select the right and physical profile. As for the dark halo, the typical value for the fluid velocity dispersion is about $100~\textrm{km}~\textrm{s}^{-1}$, which is roughly also the value of $a_{\infty} \approx 10^{-3.5}$ in unit of light speed. Thus, we solve $x=x(\bar{r})$, or equivalently $\bar{r}=\bar{r}(x)$, from \eq{a_r_eq}, and only real solution can satisfy the boundary conditions. Given the sound speed profile, we can in turn obtain the mass density profile $\rho_0(\bar{r})$.

\section{Local Mach number}\label{app:mach}
The proper speed of the fluid $v\equiv\abs{v^r}=\abs{{\rm d}r_{\rm obs}/{\rm d}t_{\rm obs}}$ measured by a ``local'' and ``stationary'' observer at $r$ with the observer's proper length ${\rm d}r_{\rm obs}(r)={\rm d}r/\sqrt{1-2M/r}$ and proper time ${\rm d}t_{\rm obs}(r)=\sqrt{1-2M/r}{\rm d}t$, thus
\[
v^r=\frac{{\rm d}r_{\rm obs}}{{\rm d}t_{\rm obs}}=\frac{{\rm d}r/{\rm d}t}{1-2M/r}=\frac{u^r/u^t}{1-2M/r}
\]
and with the normalization
\[
-1=-\left(u^t\right)^2\left(1-\frac{2M}{r}\right)+\left(u^r\right)^2\left(1-\frac{2M}{r}\right)^{-1},
\]
one can readily solve for 
\begin{equation}
v=\frac{u}{\sqrt{1-2M/r+u^2}}.
\end{equation}
Hence at $r\gg2M$, $v\simeq u\ll1$ and is subsonic; while at $r=2M$, $v\equiv1>a$, \emph{independent} of $u$, supersonic. Then we define the ``local'' Mach number
\be
\mathcal{M}\equiv\frac{v}{a}=\frac{u/a}{\sqrt{1-2M/r+u^2}}.
\ee
Note that at the sound horizon ($v_s=a_s$), $\mathcal{M}_s=v_s/a_s=1$; at the event horizon ($v_h=1$), $\mathcal{M}_h=v_h/a_h=1/a_h$.






\bibliographystyle{JHEP}
\bibliography{bondi} 


\end{document}